\documentclass[conference]{IEEEtran}
\IEEEoverridecommandlockouts
\usepackage{cite}
\usepackage{amsmath,amssymb,amsfonts}
\usepackage{algorithmic}
\usepackage{algorithm}
\usepackage{graphicx}
\usepackage{textcomp}
\usepackage{xcolor}

\def\BibTeX{{\rm B\kern-.05em{\sc i\kern-.025em b}\kern-.08em
    T\kern-.1667em\lower.7ex\hbox{E}\kern-.125emX}}
\ifCLASSOPTIONcompsoc
\usepackage[caption=false,font=normalsize,labelfon
t=sf,textfont=sf]{subfig}
\else
\usepackage[caption=false,font=footnotesize]{subfi
g}
\fi
\begin{document}

\title{Neural Networks Meet Elliptic Curve Cryptography: A Novel Approach to Secure Communication}


%

\author{
    \IEEEauthorblockN{Mina Cecilie Wøien\IEEEauthorrefmark{1}, Ferhat Ozgur Catak\IEEEauthorrefmark{1}}
    \IEEEauthorblockA{\IEEEauthorrefmark{1}\textit{Electrical Eng. and Computer Science} \\
    \textit{University of Stavanger}\\
    Rogaland, Norway \\
    mc.woien@stud.uis.no, f.ozgur.catak@uis.no}
    
    \and
    
    \IEEEauthorblockN{Murat Kuzlu}
    \IEEEauthorblockA{\textit{Batten College of Eng. and Tech.} \\
    \textit{Old Dominion University}\\
    Norfolk, VA, USA \\
    mkuzlu@odu.edu}

    \and
    
    \IEEEauthorblockN{Umit Cali}
    \IEEEauthorblockA{\textit{School of Physics, Eng. and Technology} \\
    \textit{University of York}\\
    York, UK \\
    umit.cali@york.ac.uk}
}

\maketitle

\begin{abstract}
In recent years, neural networks have been used to implement symmetric cryptographic functions for secure communications. Extending this domain, the proposed approach explores the application of asymmetric cryptography within a neural network framework to safeguard the exchange between two communicating entities, i.e., Alice and Bob, from an adversarial eavesdropper, i.e., Eve. It employs a set of five distinct cryptographic keys to examine the efficacy and robustness of communication security against eavesdropping attempts using the principles of elliptic curve cryptography. The experimental setup reveals that Alice and Bob achieve secure communication with negligible variation in security effectiveness across different curves. It is also designed to evaluate cryptographic resilience. Specifically, the loss metrics for Bob oscillate between 0 and 1 during encryption-decryption processes, indicating successful message comprehension post-encryption by Alice. The potential vulnerability with a decryption accuracy exceeds 60\%, where Eve experiences enhanced adversarial training, receiving twice the training iterations per batch compared to Alice and Bob.
\end{abstract}


\section{Introduction}
\label{sec:introduction}
Artificial Intelligence (AI) has provided significant advances in various domains, such as facilitating the integration of sophisticated analytical and predictive capabilities. The advanced capabilities of AI-based methods, such as rapid data processing and pattern recognition, make them a superior alternative to traditional methodologies, especially in scenarios requiring accurate decision-making \cite{perfectlySecure}. Cryptography is essential in securing smart grids and ensuring data privacy, authentication, and efficient communication \cite{mahmoud2015investigating}.  Public key cryptography, Elliptic Curve Cryptography (ECC), and identity-based cryptography are among the techniques used in the field of smart grids \cite{zhang2016elliptic}. The techniques prevent various attacks on devices and address resource constraints in smart meters \cite{10.18280/ijsse.130108}. Similarly, AI also has significant contribution to natural language processing (NLP), i.e., enabling the development of sophisticated text generation algorithms and improving interactive technologies such as ChatGPT, improving user engagement and service delivery in customer service applications \cite{MSMARCO}. Cryptography encompasses several methods to transform plain messages into encrypted ciphertexts, which can only be decrypted by authorized parties possessing the required key. There are two types of cryptography methods to achieve this. 


In recent years, the relationship between AI and cryptography has increased significantly along with exploring the potential of neural networks for algorithmic cryptography. Recent studies have demonstrated the ability of neural networks to design secure communication protocols using symmetric cryptography \cite{survey}. The study \cite{8590945} introduces a new approach to cryptography utilizing neural networks, i.e., neural networks automatically generate cryptographic schemes. The experimental results of the study also confirm the ability of the architecture for automatic encryption and decryption, along with the achievement of a neural symmetric cryptosystem. The other study \cite{symmetric} in Adversarial Neural Cryptography proposes a framework in which neural networks, through adversarial training, learn to encrypt and decrypt messages, effectively mimicking cryptographic functions. The main contribution of the current study is the application of asymmetric cryptography within an AI context, employing ECC to secure communications between neural network entities, herein referred to as Alice and Bob, against an eavesdropping entity, Eve. This study contributes to the burgeoning field of neural cryptography by demonstrating the feasibility of employing asymmetric cryptographic principles in neural network-based secure communication protocols.

The remainder of this paper is organized as follows. Section \ref{ch:related_work} provides the contemporary research landscape in neural cryptography. Section \ref{ch:backgorund} provides background related to cryptography and neural networks. Section \ref{ch:system_model} details the asymmetric neural cryptography model used in this study. The experimental results are presented and discussed in Section \ref{ch:results}, followed by the conclusion remarks and directions for future research in Section \ref{ch:conclusion}. The developed source code for this project is publicly available on GitHub repository\footnote{https://github.com/minawoien/Neural-Cryptography}.

\section{Related Work}\label{ch:related_work}
The integration of neural networks into the cryptography domain represents a novel intersection of AI and information security, offering innovative avenues for the development of secure communication protocols. The exploration of AI as a mechanism for implementing cryptographic functions has garnered increasing interest, particularly in enhancing security against sophisticated adversarial models. Zhou et al. \cite{security-in-new-models} provide a comprehensive examination of this emerging paradigm, focusing on applying neural networks to encryption tasks and assessing security in the presence of powerful adversaries. Their work underscores the potential of deep learning (DL) techniques to redefine traditional cryptographic practices.

The seminal work by Abadi and Andersen \cite{symmetric} marked a pivotal advance in the field, demonstrating the ability of neural networks to learn encryption and decryption processes autonomously through adversarial training. In their framework, a trio of neural networks engage in a cryptographic game in which Alice and Bob aim to secure their communication from the eavesdropping attempts by Eve, facilitated by a shared secret key. This study established the foundation for future studies on adversarial neural cryptography or Adversarial Neural Cryptography (ANC). However, the ANC model proposed by Abadi and Andersen has been investigated with regard to security concerns. Coutinho et al. \cite{perfectlySecure} critically analyzed the ANC model in terms of its security assessment. They indicated that the proposed ANC model does not adequately simulate realistic adversarial conditions predicated on Eve's difficulty in decrypting messages without access to the ciphertext. They proposed an enhanced model, i.e., Chosen-Plaintext Attack Adversarial Neural Cryptography (CPA-ANC). In this enhanced model, Eve selects one of two messages for Alice to encrypt, providing a more stringent test of the cryptographic system's security. According to the results, the CPA-ANC model demonstrates improved resilience against adversarial decryption attempts, while the original ANC model may not offer robust security. Meraouche et al. \cite{asym-neural} extended the concept of adversarial neural cryptography by incorporating asymmetric cryptographic principles based on these foundational studies. Their model introduces additional neural networks, which generate key pairs, facilitating secure communication through public and private keys.

All of these studies contribute to adversarial neural cryptography by systematically analyzing the application of asymmetric cryptographic techniques within a neural network framework. This study aims to evaluate the efficacy of neural networks in securing communications against eavesdropping attempts using ECC. 

\section{Background}\label{ch:backgorund}

\subsection{Asymmetric Cryptography}
Asymmetric cryptography is one of the widely used encryption methods, which is also known as \textit{public-key cryptography}. It uses a public key-private key pairing, i.e., a public key for encryption and a private key for decryption, and differs from symmetric cryptography using the same key both encrypts and decrypts data. 

ECC is a cryptographic approach used in asymmetric cryptography, which utilizes the principles of elliptic curve theory. It is also considered an alternative to the Rivest-Shamir-Adleman (RSA) cryptographic algorithm, which is frequently used for digital signatures. ECC provides several advantages, i.e., faster, smaller, and more efficient in generating cryptographic keys. In particular, the efficiency of the ECC makes it well-suited for environments with limited computational resources \cite{ECC}.



\subsection{Generative Adversarial Networks (GANs)}
Generative Adversarial Networks (GANs) introduce a novel approach within the realm of unsupervised learning, consisting of two competing neural network models: a generator producing synthetic data similar to training data and a discriminator evaluating the authenticity of real and generated data \cite{creswell2018generative}. Through iterative adversarial training, both networks improve their performance, with the generator producing increasingly convincing data and the discriminator becoming more adept at distinguishing between real and generated inputs. This approach has been applied to many applications such as image generation, image-to-image translation, text-to-image translation, 3D object generation, semantic-image-to-photo Translation, style transfer, and more recently, adversarial neural cryptography \cite{sarp2021wg2an, 10155107}.

\section{System model}\label{ch:system_model}
This section describes the proposed asymmetric neural cryptography model built on Pacurar's implementation of asymmetric neural cryptography\footnote{https://mathybit.github.io/adversarial-neural-crypto/}. 

\subsection{Overview}\label{sec:overview}
\begin{figure}[htbp]
\centerline{\includegraphics[width=1.0\linewidth]{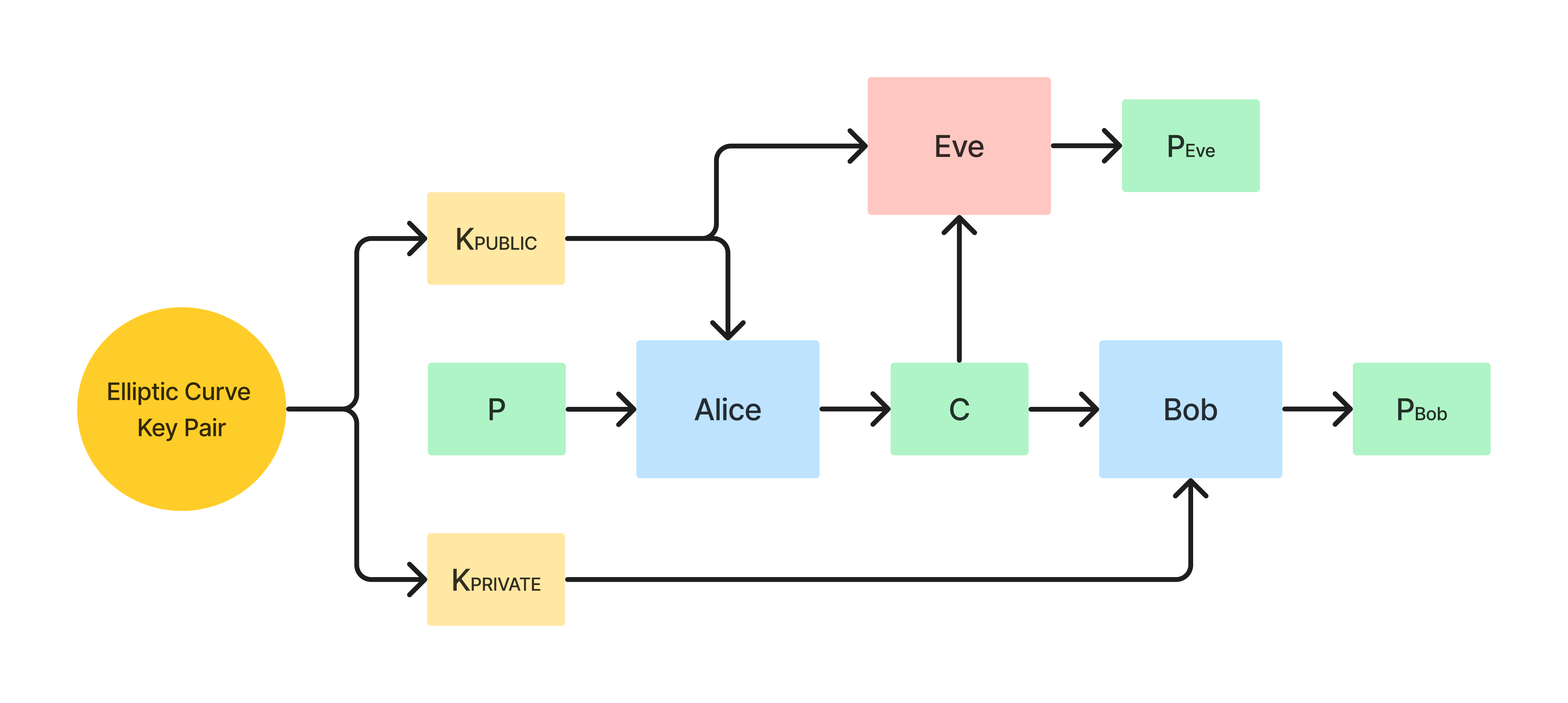}}
\caption{Asymmetric cryptosystem.}
\label{fig:system}
\end{figure}

The proposed asymmetric neural cryptography model in this study includes the deployment of neural network architectures to simulate participants in a typical cryptographic system, i.e., sender, receiver, and eavesdropper, shown in Fig.~\ref{fig:system}. The model is implemented by engaging with three separate neural networks, namely Alice, Bob, and Eve. Bob initiates the communication by generating a pair of keys, i.e., a public key, \(K_{\text{PUBLIC}}\), and a private key, \(K_{\text{PRIVATE}}\). The public key is openly shared with Alice (and accessible to Eve), while the private key remains confidential to Bob. Alice utilizes \(K_{\text{PUBLIC}}\) to encrypt her plaintext message, \(P\), producing a ciphertext, \(C\). This encrypted message is then transmitted to Bob, who employs \(K_{\text{PRIVATE}}\) to decrypt \(C\) and recover the original message \(P\). At the same time, Eve endeavors to intercept and decipher the encrypted message without having Bob's private key, representing the adversarial risk in this cryptographic scenario. The success of the model depends on the challenge of inferring \(K_{\text{PRIVATE}}\) from \(K_{\text{PUBLIC}}\). In addition, the utilization of neural networks in this context aims not only to replicate the encryption and decryption processes but also to resist eavesdropping attempts by Eve adaptively to guarantee the confidentiality of Alice and Bob's communication.

\subsection{Architecture}\label{sec:architecture}
\begin{figure}[!htbp]
\centerline{\includegraphics[width=0.9\linewidth]{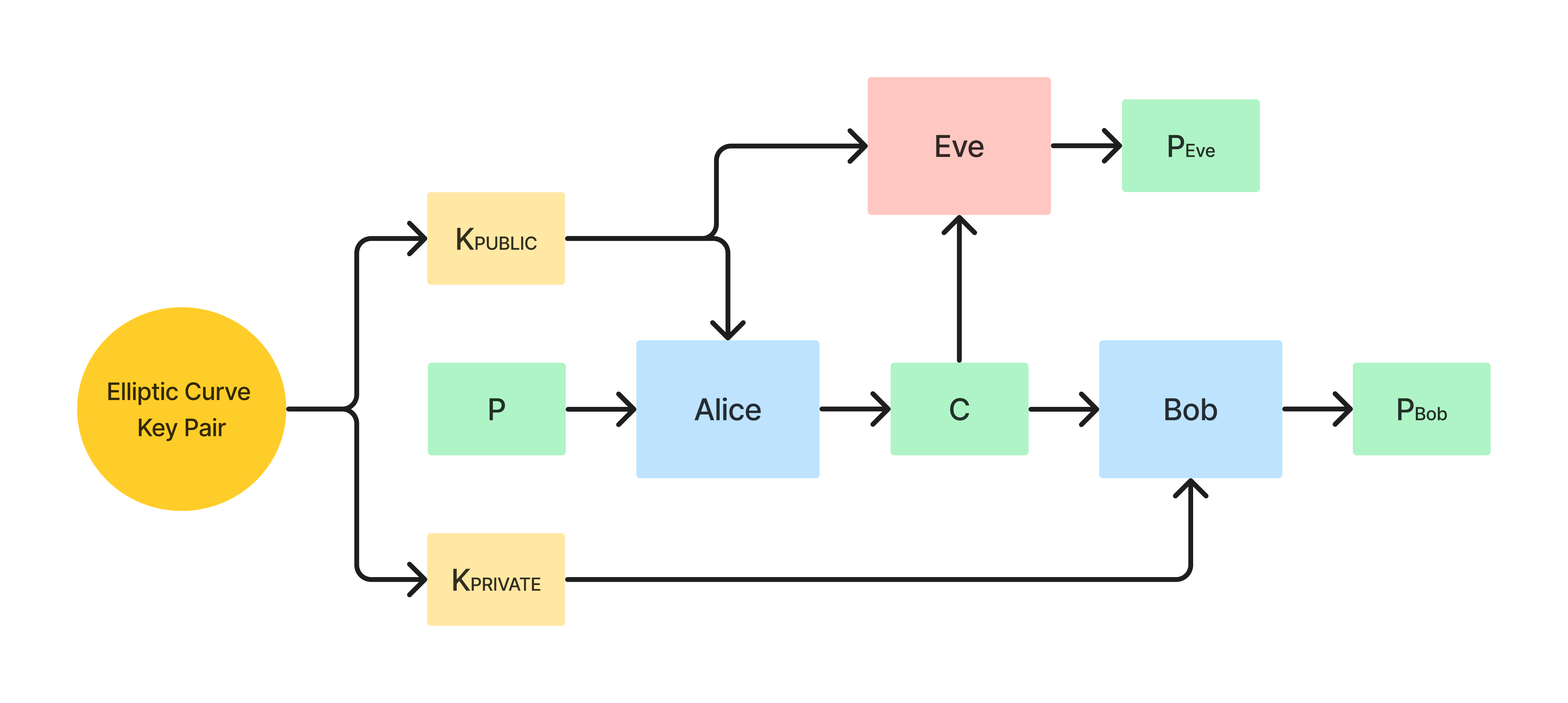}}
\caption{Overview of the system, the interaction between Alice, Bob, and Eve.}
\label{fig:system1}
\end{figure}

The architecture is adapted from generative adversarial networks (GANs) to the cryptographic concept. It consists of developing a specialized neural network architecture for secure communication using the principles of ECC \cite{ECC}. As seen in Fig.~\ref{fig:system1}, there are three primary neural network entities, i.e., Alice, Bob, and Eve, and each entity represents the sender, receiver, and eavesdropper, respectively. The most critical process in the architecture is to generate an ECC key, where Bob generates a key pair consisting of a public key, \(K_{\text{PUBLIC}}\), and a private key, \(K_{\text{PRIVATE}}\). 

\begin{figure}[!htbp]
\centerline{\includegraphics[width=0.9\linewidth]{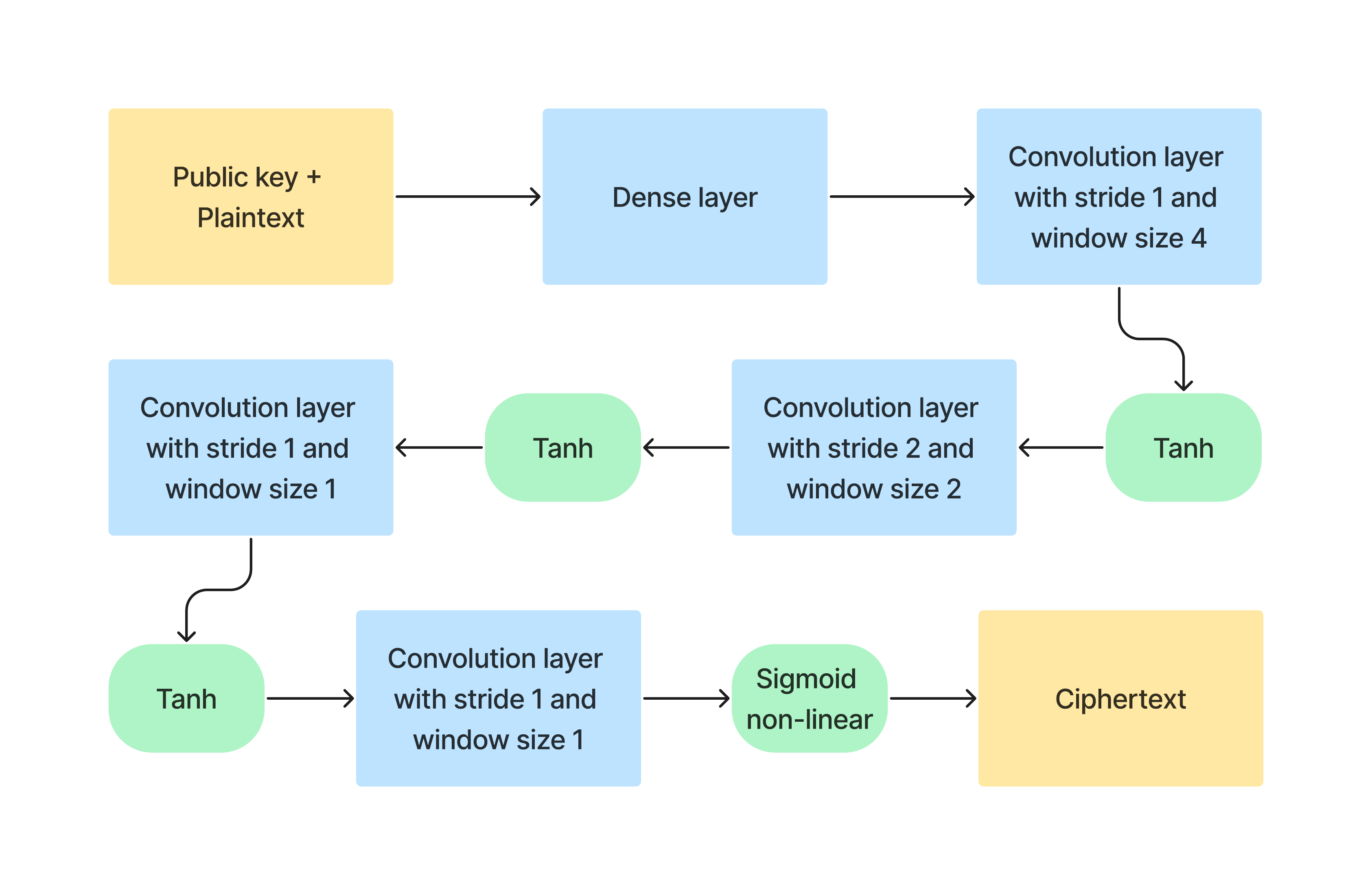}}
\caption{Encryption process flow, illustrating how Alice encrypts a plaintext message using Bob's public key.}
\label{fig:flow}
\end{figure}

Alice's network encrypts messages in plaintext., \(P\), using Bob's public key to generate ciphertext, \(C\), as shown in Fig.~\ref{fig:flow}. This encrypted message is transmitted to Bob, who utilizes his private key to decrypt the message. The neural network architectures of Alice and Bob are designed to optimize this encryption-decryption cycle, and guarantee the confidentiality and integrity of the transmitted information. Fig.~\ref{fig:system1} illustrates the overall communication system, highlighting the information flow and the interaction among neural networks.

Eve intercepts the ciphertext and attempts decryption without access to Bob's private key, i.e., an eavesdropper. Eve's network challenges the encryption process to simulate adversarial attempts to breach communication security.


The training of neural networks uses an iterative adversarial training methodology similar to that used in GANs. This approach involves alternating between training Alice and Bob to enhance their communication security and training Eve to improve her eavesdropping capabilities. The objective is to reach an equilibrium where Alice and Bob can securely communicate without Eve successfully decrypting the messages, thereby ensuring the effectiveness of the proposed asymmetric neural cryptography model.

\subsection{Training}

\begin{figure}[!htbp]
\centerline{\includegraphics[width=0.9\linewidth]{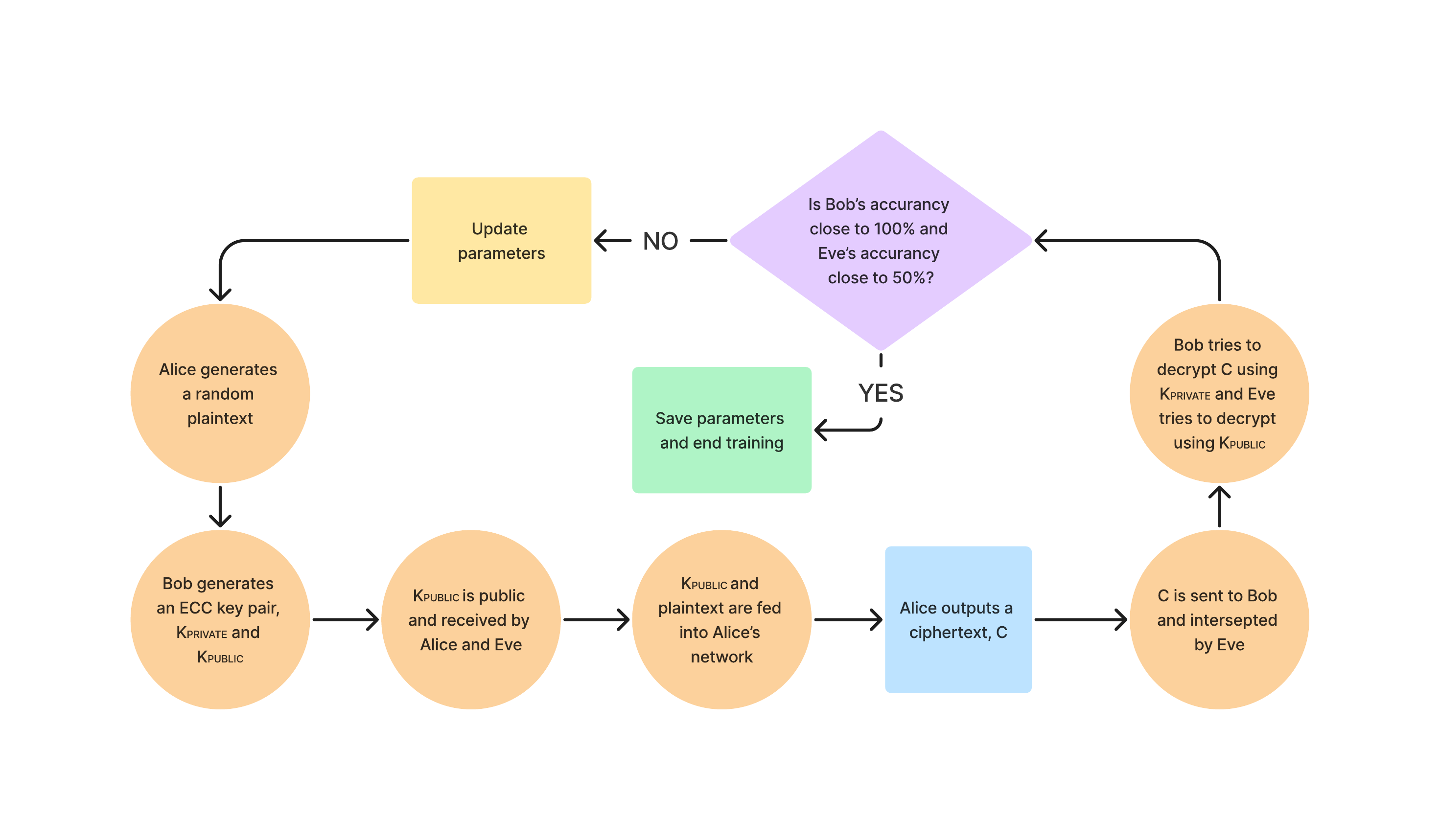}}
\caption{Training flow of the neural networks.}
\label{fig:training}
\end{figure}

The training process is illustrated in Fig.~\ref{fig:training}, Alice initiates the process by creating a plaintext message that she wants to send to Bob privately. On the other hand, Bob generates a key pair using an elliptic curve consisting of a public key and a private key. The public key is shared between Alice and Eve. Alice then encrypts the plaintext message using the public key, which produces a ciphertext. Both Bob and Eve receive this ciphertext. Bob attempts to decrypt the ciphertext using his private key, while Eve tries to decrypt it using only Bob's public key. If Bob's decrypted message matches the original plaintext with a high degree of accuracy and Eve's accuracy is significantly lower, the parameters are saved, and the training process ends. If not, the parameters are updated, and the training process continues.

During the training phase, the ABE model, Eve model, and Bob's loss are computed using the same approach as that of Pacurar's implementation. The loss of Bob signifies his proficiency in decrypting messages, while the loss of Eve indicates her ability to intercept the communication between Alice and Bob. This loss is determined by subtracting the predicted outcome - the output produced by Eve or Bob - from the target output, which is Alice's plaintext message. Bob's loss is presented in Eq. ~\ref{eq:bobLoss}.

\begin{equation}
    \ell\textsubscript{BOB} = \frac{1}{N} \sum_{i=1}^{N} \sum_{j=1}^{m_{\text{bits}}} \left| P{_{ij}} - O_{\text{ij}} \right|
    \label{eq:bobLoss}
\end{equation}
where $\ell\textsubscript{BOB}$ is the loss function, $P_{ij}$ is the plaintext, $O_{ij}$ is Bob's predicted outcome, $N$ is the batch size, and $m\textsubscript{bits}$ is the message size. Eve's loss is defined in Eq. ~\ref{eq:eveLoss}.

\begin{equation}
    \ell\textsubscript{EVE} = \frac{1}{N} \sum_{i=1}^{N} \sum_{j=1}^{m_{\text{bits}}} \left| P{_{ij}} - O_{\text{ij}} \right|
    \label{eq:eveLoss}
\end{equation}
where $\ell\textsubscript{EVE}$ is the loss function and $O_{ij}$ is Eve's predicted outcome.
The ABE loss is calculated concerning Bob's loss and Eve's loss as shown in Eq.~\ref{eq:ABE-loss}

\begin{equation}
    \ell\textsubscript{ABE} = \ell\textsubscript{BOB} + \frac{\left(\frac{m_{\text{bits}}}{2} - \ell\textsubscript{EVE}\right)^2}{\left(\frac{m_{\text{bits}}}{2}\right)^2}
    \label{eq:ABE-loss}
\end{equation}
where $\ell\textsubscript{ABE}$ is the loss function of the ABE model, $\ell\textsubscript{BOB}$ is Bob's loss function, $\ell\textsubscript{EVE}$ is Eve's loss function and $m\textsubscript{bits}$ is the message size. This loss indicates whether Alice and Bob need to improve their strategy, as it shows how well Alice and Bob work together while Eve performs poorly. If Eve is doing too well, this loss will be high.


Neural network models are trained using an epoch size of 20 and a batch size of 512. An epoch refers to the number of times that models are trained in the entire data set, while batch size signifies the number of data set rows that are used simultaneously \cite{optimizer}. The message space is defined as $2^{m_{\text{bits}}}$, where m\textsubscript{bits} denotes the length of the message of 16. The number of iterations per epoch is calculated by dividing the message space by the batch size, resulting in 2500 iterations.

During the training of the ABE model, a set of keys is generated for each message. The process entails the creation of 512 batches of messages consisting of 16 bits and 512 batches of public/private keys. The same approach is implemented for the Eve model, except that the private key is not accessible.

\begin{algorithm}[!htbp]
\scriptsize
\caption{Train the ABE model.}
\begin{algorithmic}[1]
\STATE epoch = 0
\STATE abeLoss = []
\STATE bobLoss = []
\WHILE{epoch $<$ n\_epochs}
    \FOR{iteration from 0 to n\_batches}
        \STATE Set Alice.trainable to True
        \FOR{cycle in from 0 to abecycles}
            \STATE messages = generateRandomBatchMessages()
            \STATE privKeys, pubKeys = getBatchKeyPairs()
            \STATE loss = abeModel.Train(messages, pubKeys, privKeys)
        \ENDFOR
    \STATE Append(abeLoss, loss)
    \STATE ciphertext = Alice.Predict(messages, pubKeys)
    \STATE decryptedText = Bob.Predict(ciphertext, privKeys)
    \STATE loss = calculateAverageLoss(decryptedText, message)
    \STATE Append(bobLoss, loss) 
    \ENDFOR
    \STATE epoch += 1
\ENDWHILE
\end{algorithmic}
\label{alg:abeTrain}
\end{algorithm}

The ABE model training process is made up of the training of Alice and Bob, as demonstrated in Algorithm~\ref{alg:abeTrain}. In each epoch, the ABE model and Bob's loss are calculated for each batch, and Alice's training is enabled. In this experiment, a single cycle generates 512 messages of size 16 and 512 key pairs per batch. These messages and key pairs are utilized to train the ABE model, with the model's weights being updated throughout the training process. After each training iteration, the loss value is determined\footnote{
https://keras.io/api/models/model\_training\_apis}. Upon completion of the training cycle, the combined loss of Alice's ability to encrypt messages and Bob's ability to decrypt them is recorded in \verb|abeLoss|.

Alice encrypts a batch of messages following the current training to determine Bob's decryption capability. Subsequently, Bob leverages the \verb|Predict| method to decrypt the encrypted messages received. This method is utilized in trained models to generate predictions. Bob's loss is calculated according to equation~\ref{eq:bobLoss} and stored in \verb|bobLoss|.

Algorithm~\ref{alg:eveTrain} outlines the Eve model's training process. During each epoch, the loss value of Eve is stored in the variable \verb|eveLoss|, while Alice is set to non-trainable to train Eve. The current study examined the effect of training Eve using one and two cycles to observe the advantage of allowing Eve to train twice for each batch, while Alice and Bob only train once. A batch of 512 messages containing 16 bits and a batch of 512 public keys were randomly generated for each cycle to train Eve. 

\begin{algorithm}[!htbp]
\caption{Train the EVE model.}
\scriptsize
\begin{algorithmic}[1]
\STATE epoch = 0
\STATE eveLoss = []
\WHILE{epoch $<$ n\_epochs}
    \FOR{iteration from 0 to n\_batches}
        \STATE Set Alice.trainable to False
        \FOR{cycle in from 0 to evecycles}
            \STATE messages = generateRandomBatchMessages()
            \STATE publicKeys = getBatchPublicKeys()
            \STATE loss = eveModel.Train(messages, publicKeys)
        \ENDFOR
    \STATE Append(eveLoss, loss)
    \ENDFOR
    \STATE epoch += 1
\ENDWHILE
\end{algorithmic}
\label{alg:eveTrain}
\end{algorithm}

\section{Results}\label{ch:results}
The study involved the application of five distinct prime curves with varying key sizes to generate elliptic key pairs for Bob. The secp224r1 curve was utilized with a key size of 224 bits, while the \textit{secp256k1, secp256r1, secp384r1, and secp521r1} curves were used with key sizes of 256, 384, and 521 bits, respectively. The ABE model was trained once for each batch, and the results were measured when Eve was trained once for one batch and when she was trained twice to test the security using improved adversarial training. The entire training process was repeated five times using the same elliptic curve and consistent variables for both the ABE and Eve models. Subsequently, the average results of these five training sessions were calculated to ensure a robust evaluation.

\begin{figure*}[!t]
\centering
\subfloat[Curve: secp224r1]{\includegraphics[width=0.19\linewidth]{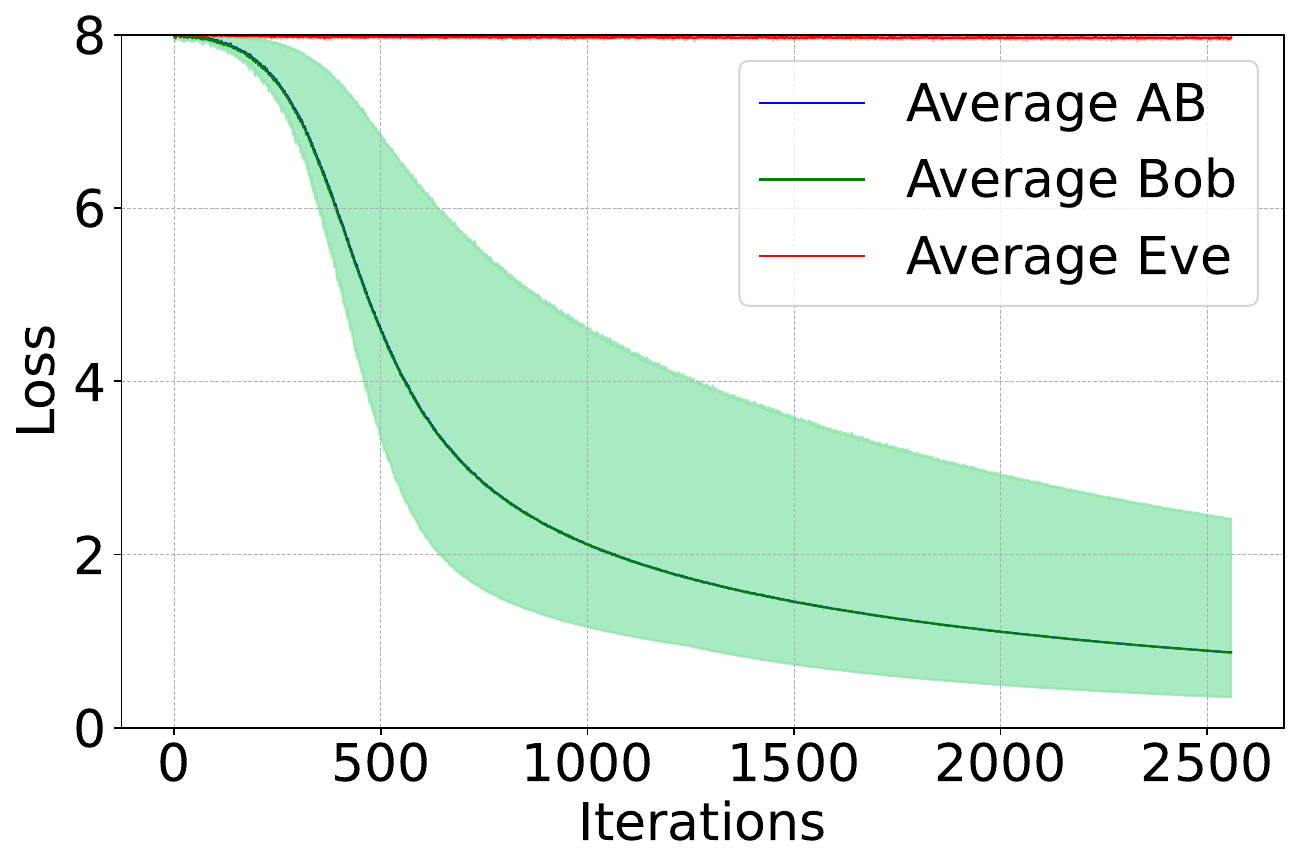}
\label{fig:224r_1}}
\hfil
\subfloat[Curve: secp256k1]{\includegraphics[width=0.19\linewidth]{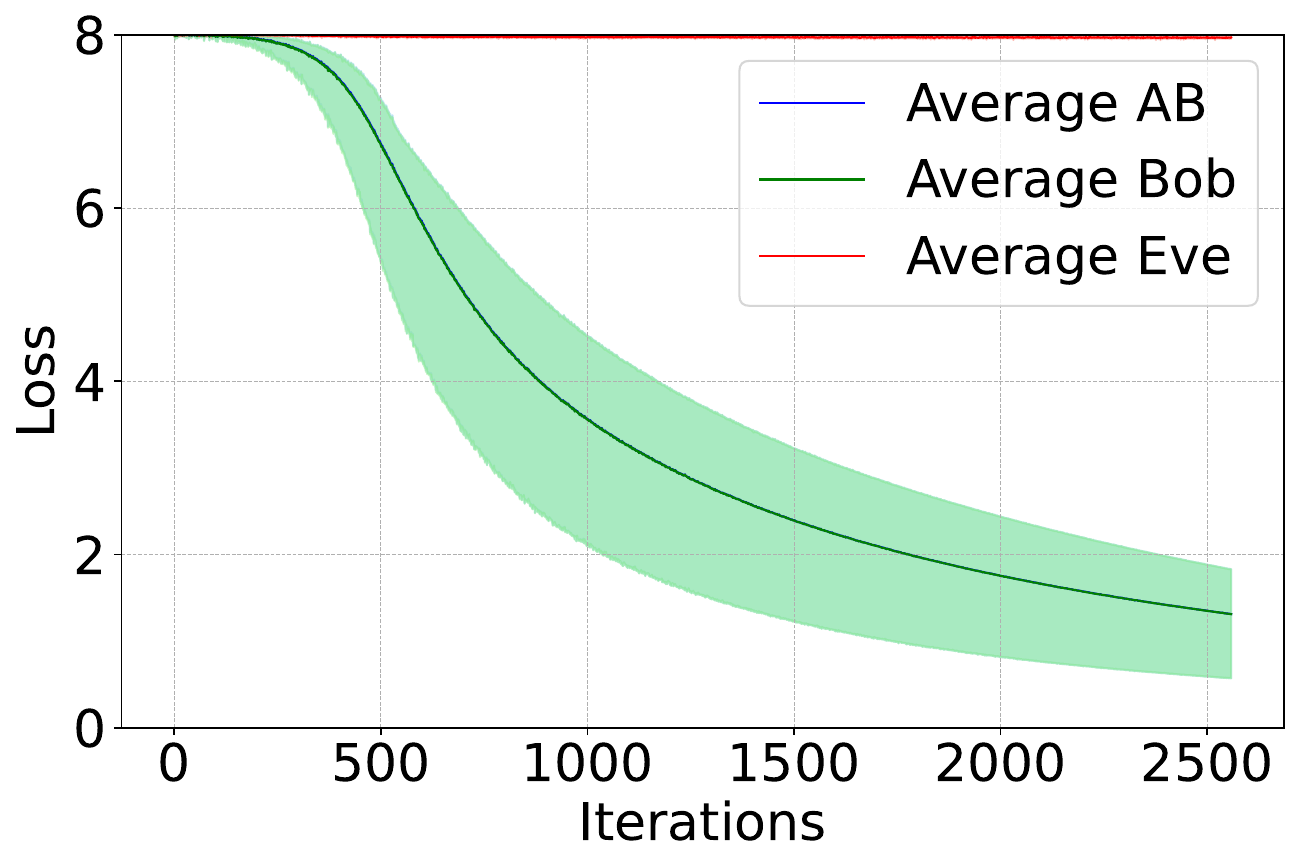}
\label{fig:256k_1}}
\hfil
\subfloat[Curve: secp256r1]{\includegraphics[width=0.19\linewidth]{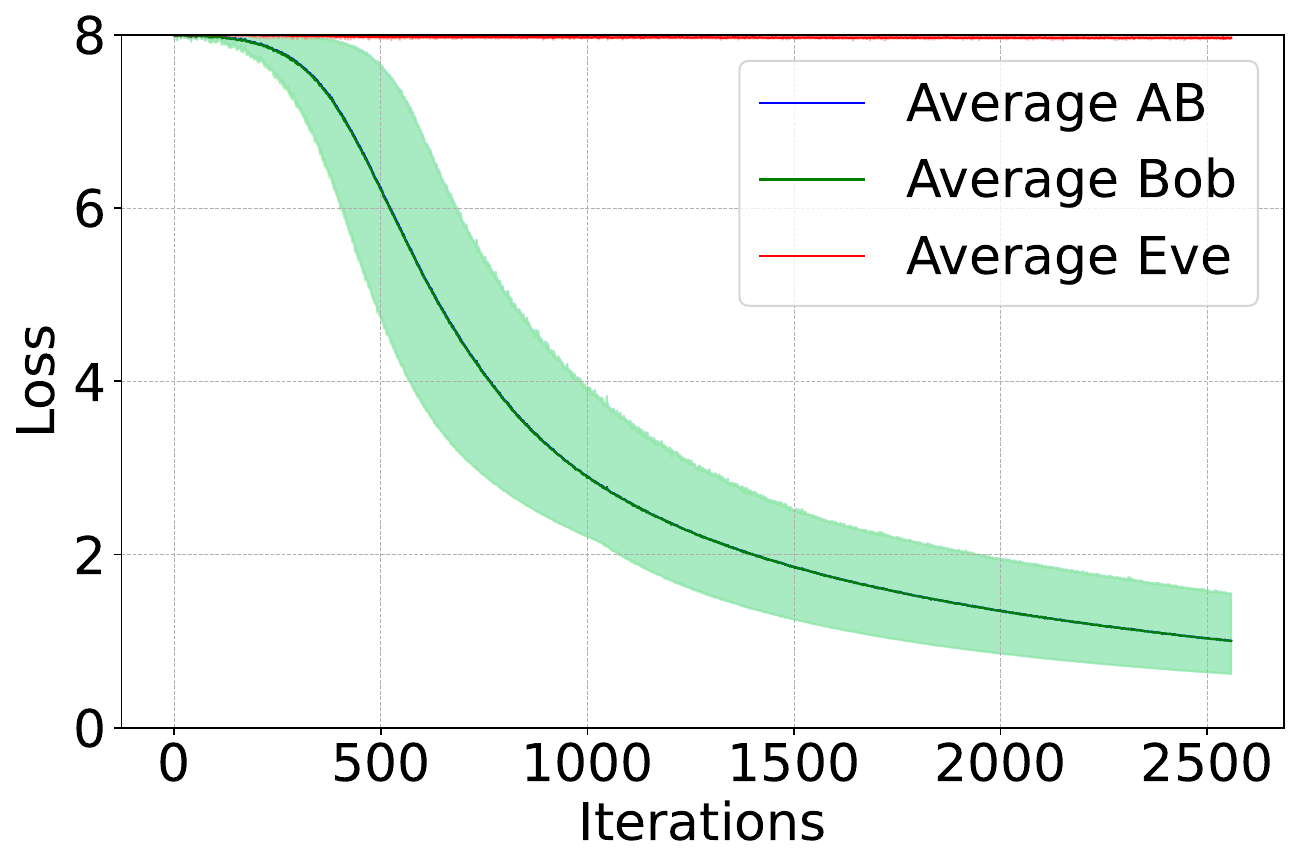}
\label{fig:256r_1}}
\hfil
\subfloat[Curve: secp384r1]{\includegraphics[width=0.19\linewidth]{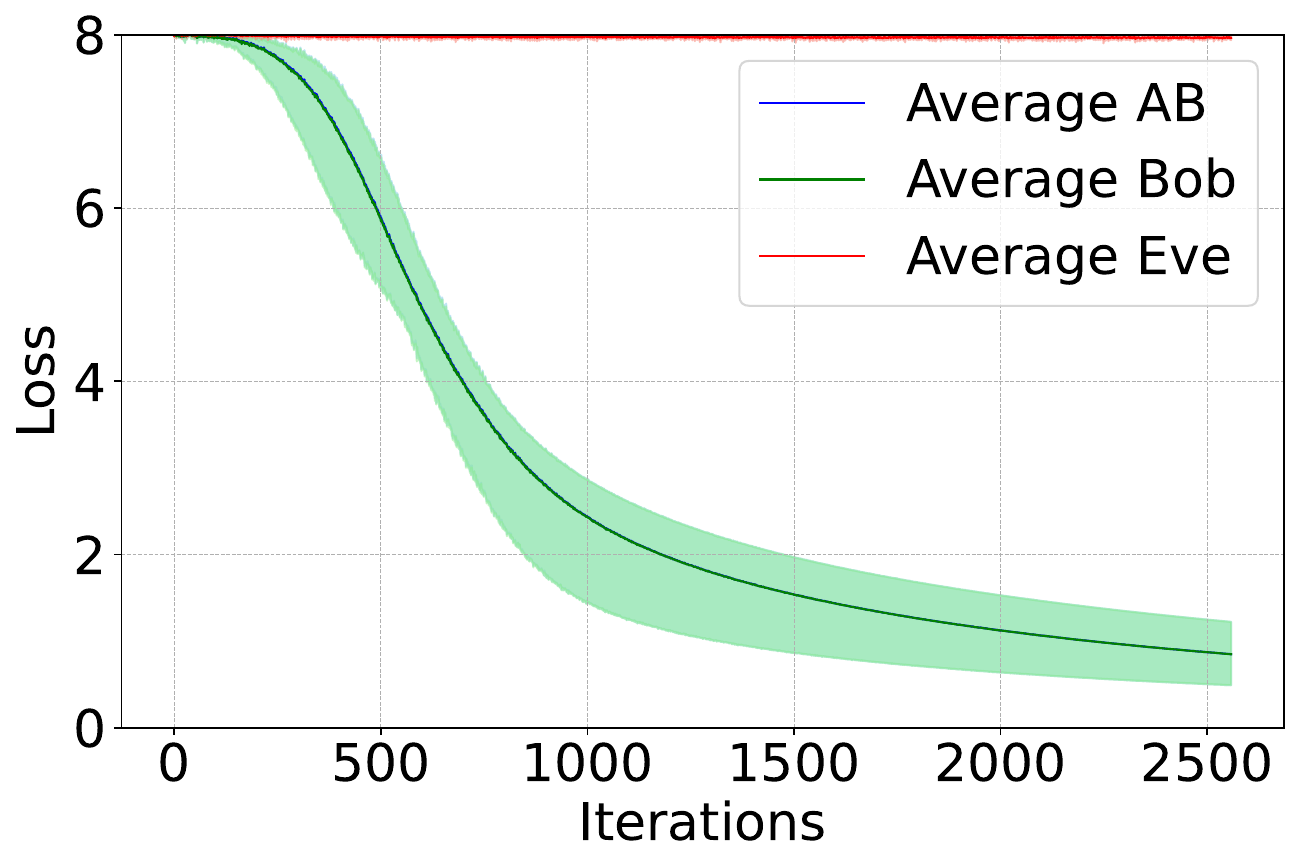}
\label{fig:384r_1}}
\hfil
\subfloat[Curve: secp521r1]{\includegraphics[width=0.19\linewidth]{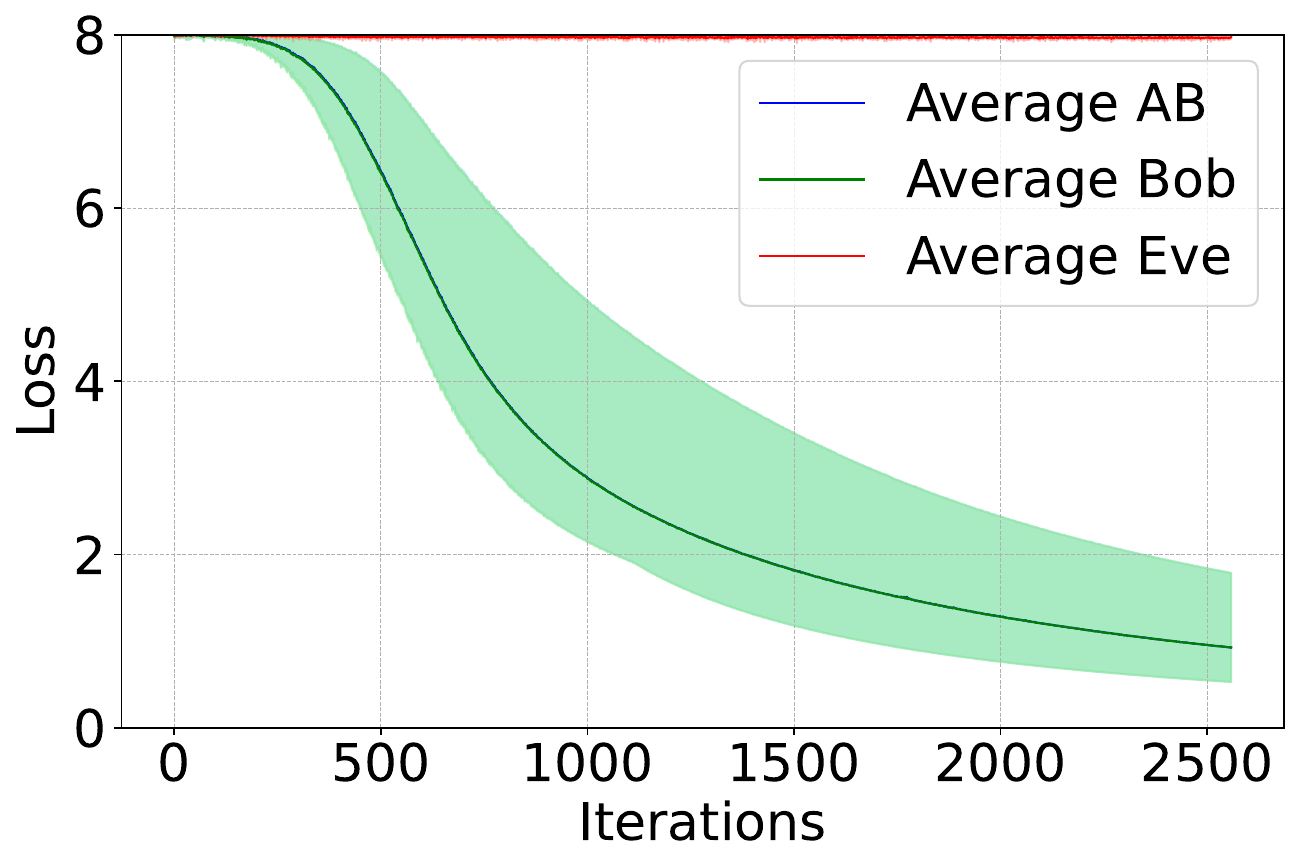}
\label{fig:521r_1}}

\caption{Loss functions for the ABE model, Bob and Eve, with different curves.}
\label{fig:1cycle}
\end{figure*}

Fig.~\ref{fig:1cycle} shows the loss functions for the ABE model, Bob and Eve, for each elliptic curve when the ABE model and the Eve model train only once for each batch. Eve has a constant loss of 8 in these figures, visualized by the red line, which is 50\% of the message size of 16 bits, indicating random guessing. The green curve represents the loss of Bob, while the blue curve represents the loss of the ABE model. These decreasing curves show that Alice and Bob are learning to protect their communication. 

The ABE model's loss values, after 2500 iterations, are shown in Table~\ref{table:loss_1}. These values are the average of five training iterations. The ABE loss indicates the relationship between how well Alice can encrypt messages and how well Bob can decrypt them, while Eve should not be able to decrypt them. At the end of the training, the ABE model has a loss value between 0 and 1, indicating the minimum loss or error in the encryption and decryption process, while Eve performs poorly. Bob also has loss values between 0 and 1, meaning that Bob's decryption of the ciphertext is close to Alice's original plaintext. Eve's loss value is approximately 8, also shown in Fig~\ref{fig:1cycle}, meaning she is not learning how to decrypt the ciphertext and is not performing better than random guessing.

\begin{table}[htbp]
\caption{Loss values for the ABE model, Bob and Eve, with different curves.}
\begin{center}
\begin{tabular}{|l|c|c|c|}
    \hline
    & \textbf{ABE loss} & \textbf{Bob loss} & \textbf{Eve loss} \\
    \hline
    \textbf{secp224r1} & 0.871 & 0.87 & 7.968 \\
    \hline
    \textbf{secp256k1} & 1.313 & 1.312 & 7.97 \\
    \hline
    \textbf{secp256r1} & 1.005 & 1.003 & 7.964 \\
    \hline
    \textbf{secp384r1} & 0.85 & 0.849 & 7.965 \\
    \hline
    \textbf{secp521r1} & 0.927 & 0.927 & 7.975 \\
    \hline
\end{tabular}
\label{table:loss_1}
\end{center}
\end{table}

After training for 2500 iterations, Alice encrypts another message to calculate Bob and Eve's decryption accuracy. The accuracy is shown in Table~\ref{table:accurancy_1cycles}. The values are calculated by adding the corrected decrypted bits and dividing them by the total number of bits. Bob decrypts the ciphertext using his private key and obtains a decryption accuracy of approximately 100\%. In four of the five different curves, Bob's accuracy is 100\%, and only one is 99.92\%, meaning that even though Bob's loss is not exactly zero at the end of the training, Bob's decryption process is highly accurate. Eve tries to decrypt the ciphertext without the private key, but only obtains a decryption accuracy of approximately 50\%. 

\begin{table}[htbp]
\scriptsize
\caption{Decryption accuracy for Bob and Eve with different curves.}
\begin{center}
\begin{tabular}{|l|c|c|}
    \hline
    &  \textbf{Bob} & \textbf{Eve} \\
    \hline
    \textbf{secp224r1} & 100\% & 51.19\% \\
    \hline
    \textbf{secp256k1} &  100\% & 55.74\% \\
    \hline
    \textbf{secp256r1} & 99.92\% & 53.51\% \\
    \hline
    \textbf{secp384r1} & 100\% & 51.914\% \\
    \hline
    \textbf{secp521r1} & 100\% & 51.29\% \\
    \hline
\end{tabular}
\label{table:accurancy_1cycles}
\end{center}
\end{table}

\begin{figure*}[!t]
\centering
\subfloat[Curve: secp224r1]{\includegraphics[width=0.19\linewidth]{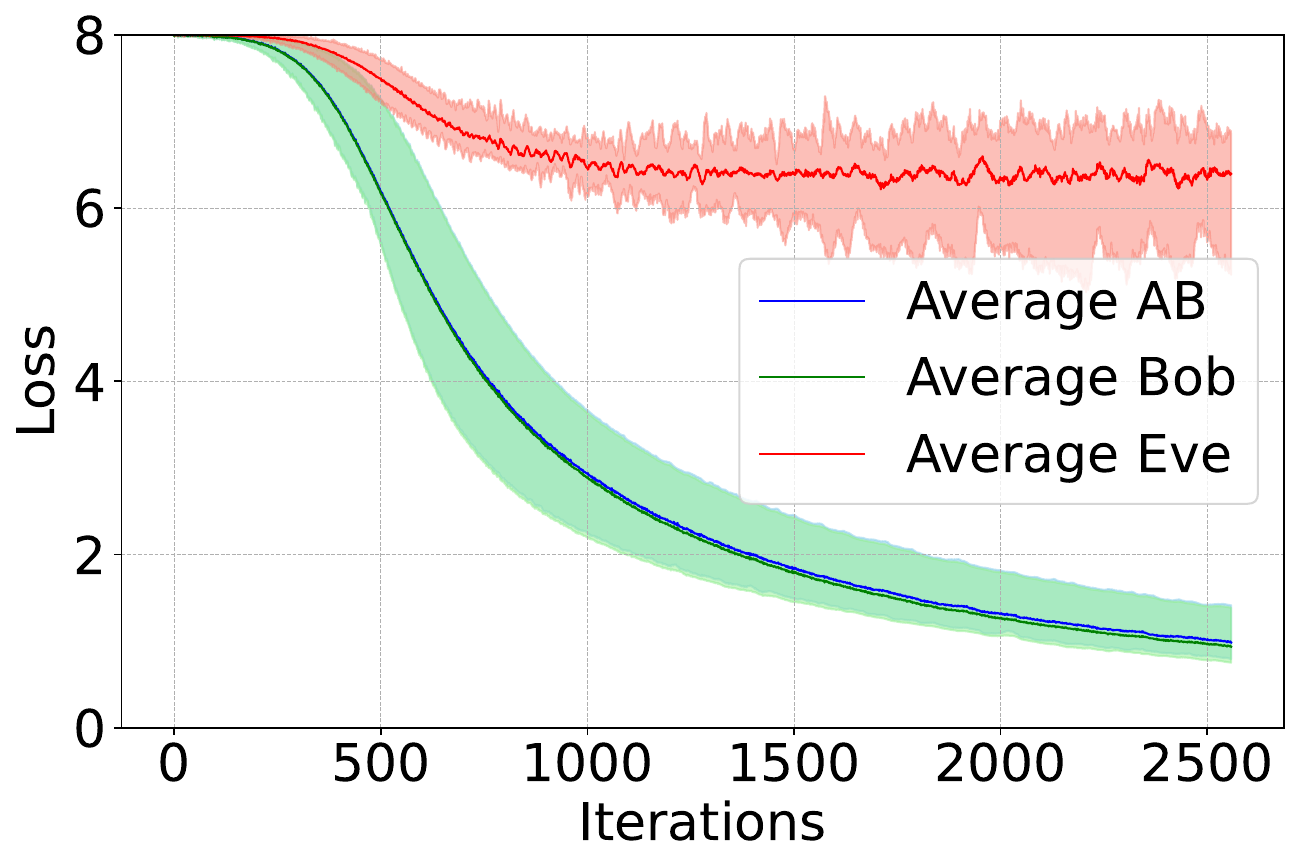}
\label{fig:224r_2}}
\hfil
\subfloat[Curve: secp256k1]{\includegraphics[width=0.19\linewidth]{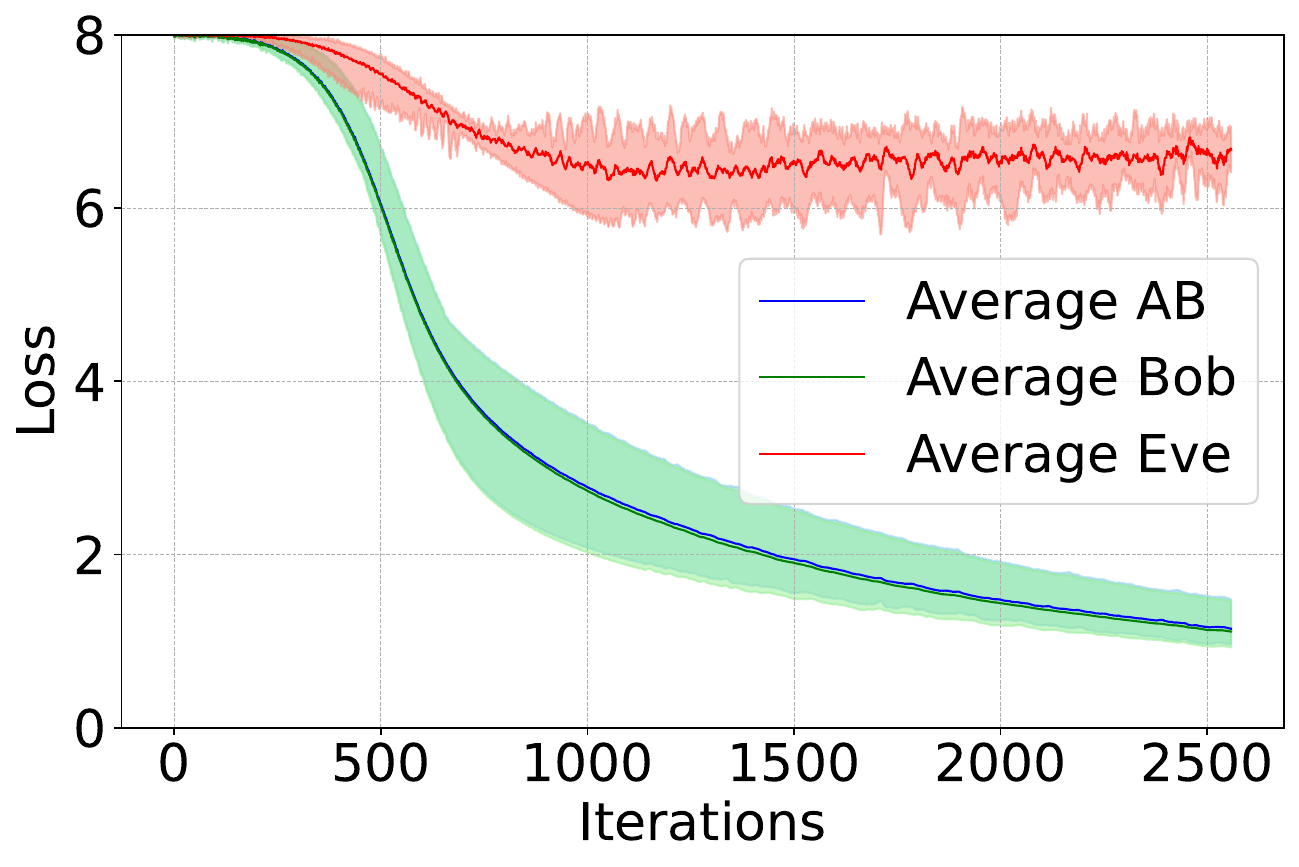}
\label{fig:256k_2}}
\hfil
\subfloat[Curve: secp256r1]{\includegraphics[width=0.19\linewidth]{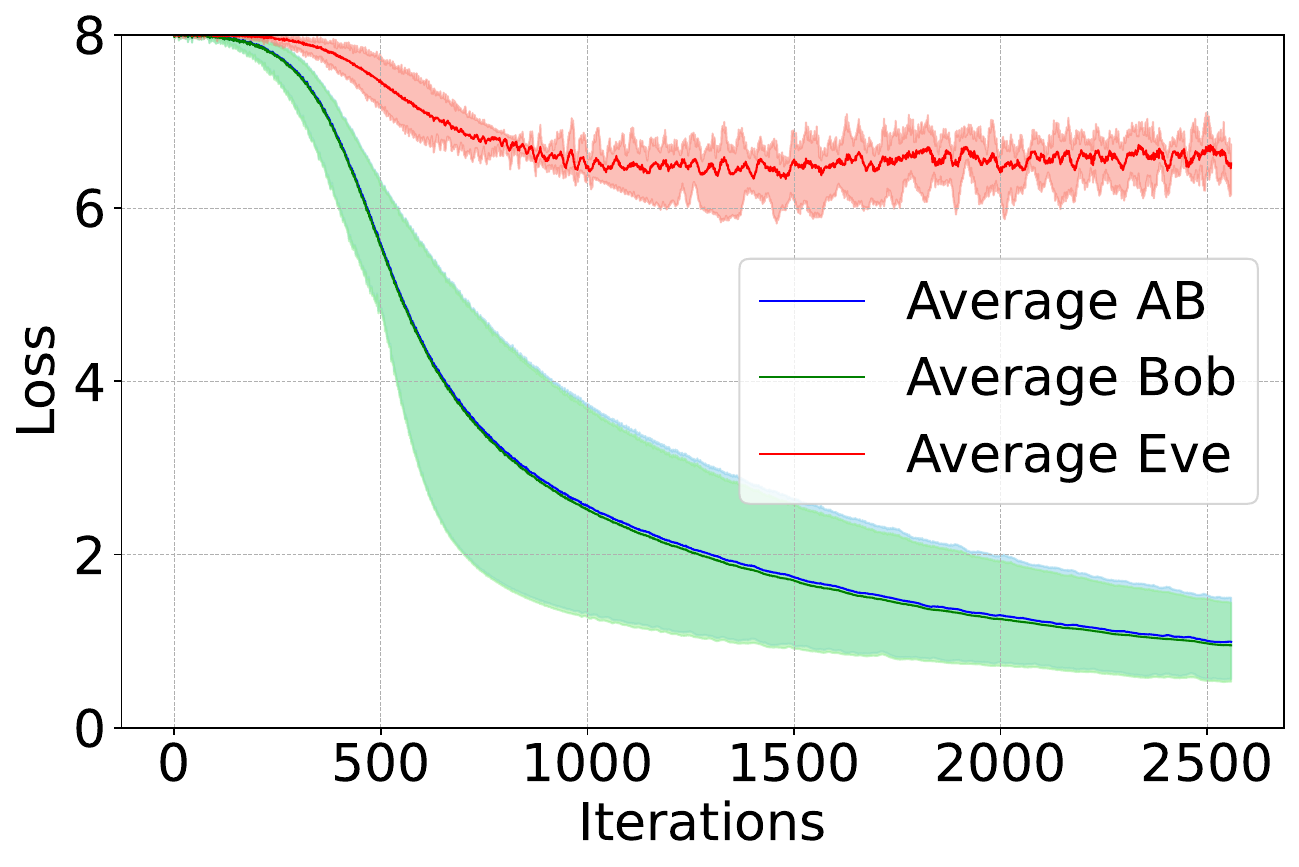}
\label{fig:256r_2}}
\hfil
\subfloat[Curve: secp384r1]{\includegraphics[width=0.19\linewidth]{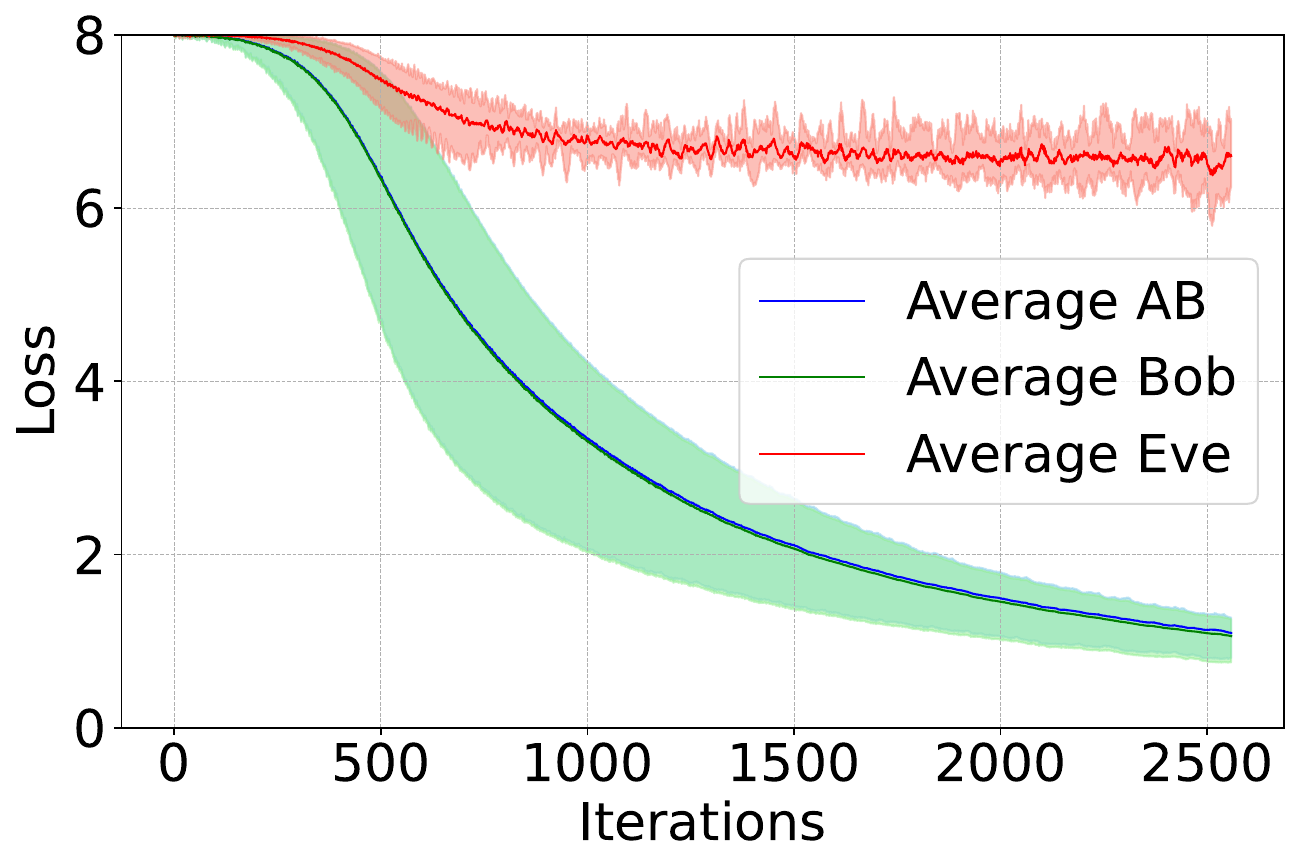}
\label{fig:384r_2}}
\hfil
\subfloat[Curve: secp521r1]{\includegraphics[width=0.19\linewidth]{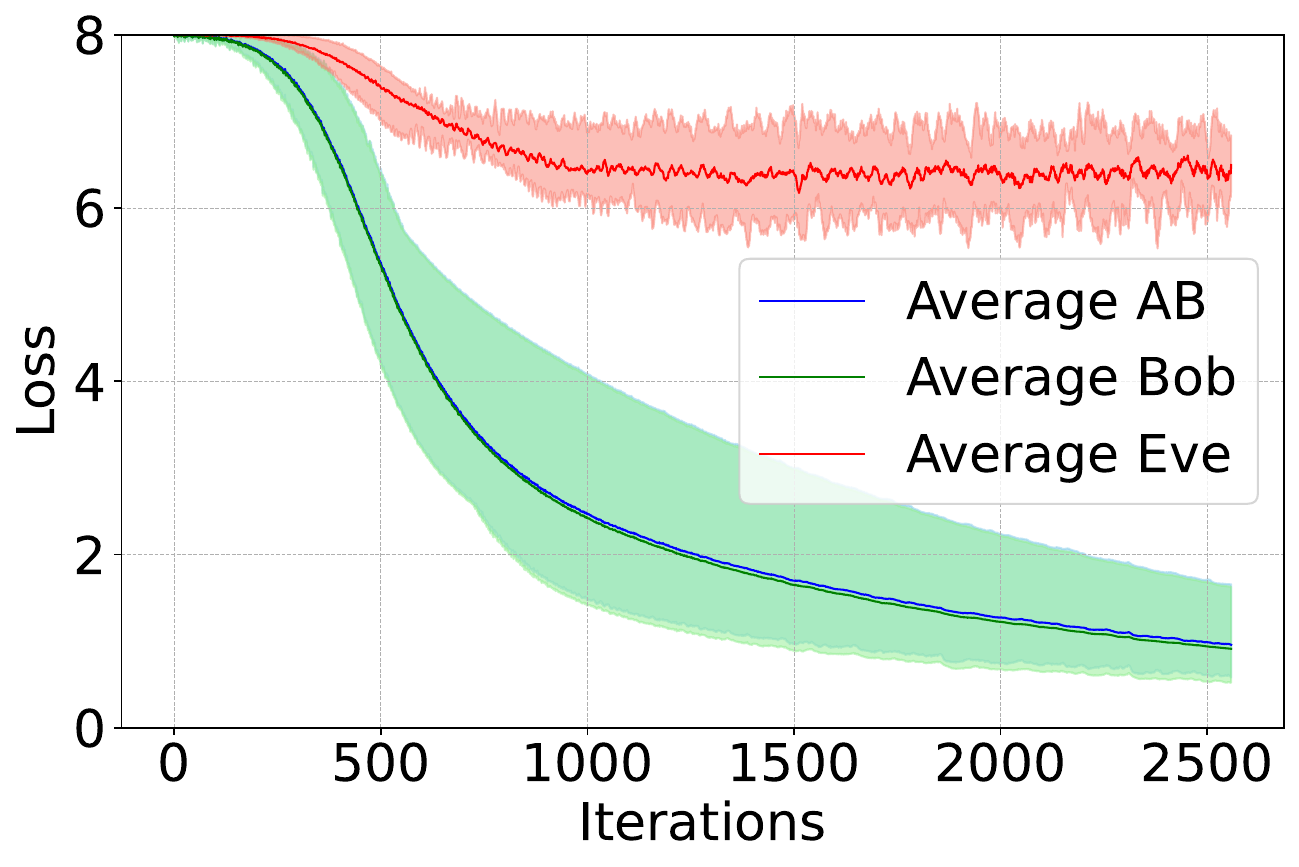}
\label{fig:521r_2}}
\caption{Loss functions for the ABE model, Bob and Eve, with different curves when Eve trains with two cycles.}
\label{fig:2cycles}
\end{figure*}

These results show that Alice and Bob can secure their communication from an eavesdropping neural network when the neural network models train once for each batch. Bob can decrypt Alice's message using the correct key, while Eve can not make sense of the message without the correct key. To further test the security, the encryption system is tested with improved adversarial training, where Eve has the advantage of training twice for each batch. At the same time, Alice and Bob only trained once. Fig.~\ref{fig:2cycles} shows the loss functions for the ABE model, Bob, and Eve for this scenario with five training iterations for the five elliptic curves. In these figures, all loss functions begin to decrease at approximately 250 iterations. The loss of the ABE model and Bob continues to decrease with the iterations, while Eve's loss function decreases for the first 1000 iterations before the loss varies between 6 and 7. This indicates that Eve is learning more because she has the advantage of training twice for each batch. In these figures, the loss function of the ABE model is slightly higher than Bob's loss function due to Eve's improvement in performance.

Table~\ref{table:loss_2} shows the loss values of the ABE model, Bob and Eve after training for 2500 iterations. The ABE loss and Bob's loss are between 0 and 1, as when Eve trained with only one cycle for each batch, as shown in Table~\ref{table:loss_1}. However, the ABE loss is slightly higher than Bob's loss, compared with the results in Table~\ref{table:loss_1}, where the ABE loss and Bob's loss were almost equal, as a result of Eve's improvement in performance. Bob's loss is nearly the same as in the previous result, meaning he is performing as well now. Eve's loss values now range from about 6.397 to 6.674 when she trains with two iterations for each batch. This is better than random guessing, which means that she correctly decrypts bits for more than 50\% of the message. Although she is not close to correctly decrypting the entire message, she has improved his decryption process compared to her attempt shown in Table~\ref{table:loss_1}.

\begin{table}[htbp]
\caption{Loss values for the ABE model, Bob and Eve.}
\begin{center}
\begin{tabular}{|l|c|c|c|}
    \hline
    & \textbf{ABE loss} & \textbf{Bob loss} & \textbf{Eve loss} \\
    \hline
    \textbf{secp224r1} & 0.985 & 0.935 & 6.396 \\
    \hline
    \textbf{secp256k1} & 1.143 & 1.111 & 6.674 \\
    \hline
    \textbf{secp256r1} & 0.992 & 0.952 & 6.514 \\
    \hline
    \textbf{secp384r1} & 1.094 & 1.058 & 6.602\\
    \hline
    \textbf{secp521r1} & 0.958 & 0.912 & 6.497 \\
    \hline
\end{tabular}
\label{table:loss_2}
\end{center}
\end{table}

Table~\ref{table:accurancy_2cycles} shows Bob and Eve's decryption accuracy. Bob has a decryption accuracy of approximately 100\%, which is the same as before, as shown in Table~\ref{table:accurancy_1cycles}, which means that Eve's improvement in performance does not affect Bob's ability to decrypt messages. However, Eve's decryption accuracy has increased from ranging between 51\% and 55\% using one cycle to between 61\% and 65\% when she trains with two cycles. It is still far from perfect accuracy, as Bob achieves using the correct key. 

\begin{table}[htbp]
\scriptsize
\caption{Decryption accuracy for Bob and Eve with different curves when Eve trains with two cycles.}
\begin{center}
\begin{tabular}{|l|c|c|}
    \hline
    &  \textbf{Bob} & \textbf{Eve} \\
    \hline
    \textbf{secp224r1} & 99.673\% & 65.256\% \\
    \hline
    \textbf{secp256k1} &  100\% & 61.33\% \\
    \hline
    \textbf{secp256r1} & 100\% & 63.83\% \\
    \hline
    \textbf{secp384r1} & 100\% & 64.614\% \\
    \hline
    \textbf{secp521r1} &99.998\% & 65.33\% \\
    \hline
\end{tabular}
\label{table:accurancy_2cycles}
\end{center}
\end{table}

\section{Conclusion}\label{ch:conclusion}
This investigation into asymmetric neural cryptography has clarified the potential of integrating ECC within AI models to strengthen secure communication channels against adversarial eavesdropping. Through an experimental protocol using a variety of ECC curves, it is quantitatively demonstrated that the proposed neural network models, representing the communicative entities Alice and Bob, can achieve a high degree of secure message exchange with minimal variation in security efficacy across different cryptographic curves. The results indicate a robust encryption-decryption mechanism capable of maintaining message integrity, as evidenced by Bob's loss metrics oscillating between 0 and 1, signifying Alice's near-perfect message reconstruction post-encryption.

The introduction of enhanced adversarial training, in which the eavesdropper model Eve underwent training iterations at double the frequency of Alice and Bob, unveiled a nuanced vulnerability within the system. Eve's decryption accuracy, surpassing 60\% under these conditions, heralds a potential security risk, albeit slight, under scenarios of advanced adversarial intervention. This finding underscores the necessity for continuous refinement of the neural cryptography model, particularly to optimize its resilience against increasingly sophisticated eavesdropping techniques.

In light of these findings, the future trajectory of this research is poised to explore the implementation of CPA methodologies. Such approaches will evaluate the cryptographic system's security, aiming to bolster its defenses and mitigate the vulnerabilities highlighted by the enhanced adversarial training paradigm. Furthermore, extending the model to incorporate dynamic mechanisms of key generation and exchange could further enhance its security framework, aligning with the evolving landscape of digital communication threats.


\bibliographystyle{IEEEtran}
\bibliography{IEEEabrv, references}

\end{document}